\documentclass[10pt,conference]{IEEEtran}
\IEEEoverridecommandlockouts

\usepackage{cite}
\usepackage{amsmath,amssymb,amsfonts}
\usepackage{booktabs}
\usepackage{algorithmic}
\usepackage{graphicx}
\DeclareGraphicsExtensions{.pdf,.png,.eps}
\usepackage{textcomp}
\usepackage{xcolor}
\usepackage{listings}
\usepackage[T1]{fontenc}
\usepackage{xspace}
\usepackage{multicol,lipsum}
\usepackage{multirow}
\usepackage{lipsum}

\definecolor{codegreen}{rgb}{0,0.6,0}
\definecolor{codegray}{rgb}{0.5,0.5,0.5}
\definecolor{codepurple}{rgb}{0.58,0,0.82}
\definecolor{backcolour}{rgb}{0.95,0.95,0.92}

\lstdefinestyle{codestyle}{
    backgroundcolor=\color{backcolour},   
    commentstyle=\color{codegreen},
    keywordstyle=\color{magenta},
    numberstyle=\tiny\color{codegray},
    stringstyle=\color{codepurple},
    basicstyle=\ttfamily\footnotesize,
    breakatwhitespace=false,         
    breaklines=true,                 
    captionpos=b,                    
    keepspaces=true,                 
    numbers=left,                    
    numbersep=5pt,                  
    showspaces=false,                
    showstringspaces=false,
    showtabs=false,                  
    tabsize=2
}

\def\BibTeX{{\rm B\kern-.05em{\sc i\kern-.025em b}\kern-.08em
    T\kern-.1667em\lower.7ex\hbox{E}\kern-.125emX}}
\definecolor{comment}{RGB}{0, 127, 0}

\begin{document}

\title{A Purpose-oriented Study on Open-source Software Commits and Their Impacts on Software Quality}

\author{
\IEEEauthorblockN{Jincheng He}
\IEEEauthorblockA{\textit{Department of Computer Science} \\
\textit{University}\\
jinchenh@usc.edu}
\and
\IEEEauthorblockN{Zhongheng He
}
\IEEEauthorblockA{\textit{Department of Computer Science} \\
\textit{University}\\
hezhongh@usc.edu}
}

\maketitle

\begin{abstract}
Developing software with the source code open to the public is very common; however, similar to its closed counterpart, open-source has quality problems, which cause functional failures, such as program breakdowns, and non-functional, such as long response times.
Previous researchers have revealed when, where, how and what developers contribute to projects and how these aspects impact software quality. 
However, there has been little work on how different categories of commits impact software quality.
To improve open-source software, we conduct this research to categorize commits, train prediction models to automate the classification, and investigate how commit quality is impacted by commits of different purposes.
By identifying these impacts, we will establish a new set of guidelines for committing changes that will improve the quality.
\end{abstract}

\section{Introduction}
\label{sec:introduction}

Open-source software development, being a popular way of developing and releasing new versions of software, not only makes the communication more efficient between remote developers, but also provides a large amount of data for researchers.
In this research, we will start by introducing the context of this research, including open source software, version control systems and different aspects of software repository mining that should be considered.
These aspects include commit impacts, purposes, commit messages, and software quality.
Based on these aspects, we design our research to first, manually categorize developer contributions, analyze how they impact software quality, automate the classification, and finally, come up with guidelines for developers to achieve higher quality.  

\subsection{Open Source Software and Version Control Systems}
Using open-source repositories has long been a common way to develop software.
Some of these projects are on an industrial scale.
As the scale has grown far beyond the level that an individual can control and manage, how to efficiently conduct quality control and project management is critical.

Most industrial-scale software is developed by iterative contributions from project teams, through ICSM \cite{icsm2014}, Agile \cite{agile2004}, DevOps \cite{devops2016} or other process models. 
In the iterations, version control systems, such as Git and SVN, play a critical role by enabling and facilitating concurrent contributions from developers.
Each revision, or commitment (hereafter ``commit''), contains diffs which are the lines developers change.


These changes can be made by developers from different areas of the world, at different times, have different purposes and have different impacts on the software \cite{qrs2020}, be they negative or positive. 
Thus, it is necessary to investigate how these differences influence software quality, and thus to better control the quality during the development and maintenance phases. 

Focusing on the different purposes of commits, we investigate how different types of commits impact software quality and propose guidelines for improving.

\subsection{Impact of Commits on Software Quality}
In projects, some commits impact software quality more than others.
For example, commits that change core modules, which modify system functionalities are more impactful than those that contain only a few lines of documentation fixes.

The level of impact can be defined in various ways to specify what to investigate. For example, in previous studies, researchers have defined impactful commits by whether they are in the core module \cite{pooyan_esem, pooyan_qrs}.
We believe that the more critical the commits are, the earlier they should be taken care of, in the sense of quality control and management.

\subsection{Purposes of Commits}
While the levels of impact differ, the commits also vary in their purposes.
For example, some commits merely add a few lines of documentation or comments to code while others refactor the entire code structure, make module-level modifications, or introduce a new feature with thousands of lines of code.

Furthermore, some commits may have multiple purposes while others have one for each.
It is common for developers to upload single-purpose commits. 
However, in commits where developers refactor code, add new dependencies, or apply minor fixes, the commits tend to grow beyond their intended task.
In this case, commits become multi-purpose, and it has been shown in a previous study \cite{qrs2020} that multi-purpose commits have negative impacts on software quality, compared to single-purpose ones.
In addition, it has been shown that some types of commits, such as ``feature add'', are more likely to have negative impacts on software quality.
Thus, it is important to investigate how different types of commits impact various aspects of quality and how they are related to the other metadata of software to create new guidelines for developers.
And such guidelines will ultimately help them improve the quality.

To achieve this, we review previous works that categorize commits and find that some of them have produced taxonomies for commit purposes \cite{Hindle_cate,alali_2008,Dragan,Swanson, Mauczka, Hindle_auto,qrs2020} which we evaluate, adopt, refine, and apply to automate the classification.

\subsection{Commit Message}
To categorize commits, we need to analyze project code and metadata, and one critical piece of the project metadata is the commit message. 
When developers push changes to online repositories, it is a common practice to add commit messages to explain their changes.

These messages provide important clues for understanding the purposes of those commits. 
As a result, it is important to analyze commit messages to help understand the purposes of the commits and categorize them.

\subsection{Software Quality}
The ultimate goal of the research is to improve software quality.
Thus, defining assessment guidelines for software quality is one of the most important issues of this research.

Software quality is evaluated using different tools depending on one's purpose. 
For example, COCOMO and COCOMO II \cite{cocomo1995} evaluate software with respect to their cost. 
PMD \footnote{https://pmd.github.io/}, SonarQube \footnote{https://www.sonarqube.org/} and FindBugs \footnote{http://findbugs.sourceforge.net/} define metrics based on software metadata and algorithms to evaluate security, vulnerability and bugs. 
CAST software \footnote{https://www.castsoftware.com} provides architecture evaluations in addition to other metrics.

In this research, we use metrics provided by SonarQube, PMD and FindBugs while also investigating the compilability of commits and use it as a metric to evaluate software quality. 
We consider it as a fundamental aspect of the quality, since a software revision is supposed to be compilable.


\subsection{Contribution}
In this paper, we review commit messages and code changes to categorize commits and refine the taxonomy.
To automate classification, we use commit messages, meta-data and metrics to train a prediction model, and improve its accuracy.
Using the taxonomy, we also analyze the relationship between the commit purposes and the software metrics, investigate how different types of commits may impact software quality, and finally come up with guidelines for developers to improve software quality. 

The main contributions of this paper consist of:
\begin{itemize}
    \item A data set that consists of 1914 commits, categorized by commit purposes.
    \item Findings on the relationship between commit purposes and software quality.
    \item A prediction model to automate commits classification.
    \item Refined the categorization to reduce ambiguity.
    \item Guidelines for developers to achieve higher software quality.
\end{itemize}

The remainder of the paper is organized as follows:
\begin{itemize}
    \item Section \ref{sec:related_works} summarizes the related works in categorizing commits, analyze software quality, and automated classification. 
    \item Section \ref{sec:questions} explains our research questions.
    \item Section \ref{sec:data} discusses our data set, its source, and how we manipulate it.
    \item Section \ref{sec:approach} illustrates how we analyze commits, train prediction model, and validate our works.
    \item Section \ref{sec:results} presents the results of our analysis.
    \item Section \ref{sec:threats} focuses on the threats to validity of our work and Section \ref{sec:conclusion} concludes this paper.
\end{itemize}

\section{Related Works}
\label{sec:related_works}

Before conducting this research, we reviewed related works on categorizing commits, modeling commit messages, and evaluating software quality.

\subsection{Commit Change Types}
The core of this research is purpose-oriented commit analysis.
The motivation comes from a previous study \cite{qrs2020} which has shown there is a statistically significant relation between the change types and the software quality by analyzing the compilability.
Thus, the first and most critical problem we address is the establishment of a generally applicable categorization of commits. 

Previous works primarily characterize commits by commit size, commit messages, and other project meta-data. 
For example, Purushothaman et al. \cite{purushothaman2004towards} propose a categorization based on whether a commit adds or deletes lines of code while Alali et al.\cite{alali_2008} examine nine open-source software systems to characterize commit properties by size --- lines of code, file count, number of code blocks, and extracted information from commit messages.
Arafat et al. \cite{arafat2009commit} and Hattori et al. \cite{hattori2008nature} also categorize commits by their sizes.
In addition, Dragan et al. \cite{Dragan} categorize commits using their method stereotypes.

On the other hand, some other studies\cite{purushothaman2004towards,Hindle_cate,qrs2020} adopt and refine the categories of maintenance tasks to categorize the commits, which is the approach we adopt.

In this research, we conduct a purpose-oriented categorization for commits based on previously-established categorizations for maintenance tasks.
This taxonomy has evolved over years.
Initially, Swanson \cite{Swanson} introduced maintenance task categories by dividing the work from developers into adaptive, corrective, and perfective.
Purushothaman et al. \cite{purushothaman2004towards} later added one more category, ``inspection,'' in addition to previous three, followed by Wang et al. whose work proposes a categorization, also based on the purpose of commits.

To determine the purposes, it has been common for researchers use commit messages. 
For example, Kaur et al. \cite{kaur_2018} have proposed a taxonomy based on commit messages. 
It consists of ``bug repair,'' ``feature addition,'' and ``general''.
However, using commit messages alone is not always sufficient to precisely categorize commits. 
In large commits with thousands of lines of code changes, developers usually only report major changes.

Hindle et al. \cite{Hindle_cate}, instead, review not only commit messages but code changes to categorize large commits and create sub-categories for Swanson's taxonomy.
They map the categories to the taxonomy of Mauczka et al. \cite{Mauczka} and apply them to automate classification \cite{Hindle_auto}. 
In a recent study, Jincheng et al. \cite{qrs2020} refined this taxonomy by reducing ambiguity.
Based on these results and methodology, we propose a further refinement of the taxonomy presented by Jincheng et al., adapted to reduce ambiguity of categories, especially the most confusing category, ``maintenance''.

\subsection{Automated Tagging}
An important application for which we establish the categorization is to automatically tag the commits based on their changes types.
Only if we succeed in tagging the commits efficiently and accurately will we be able to provide potential risk evaluations for them. 
In this way, it will be possible to integrate our work into existing software development tools.
Previous studies have created prediction models to achieve this based on their own categorizations.
For example, Hindle et al. \cite{Hindle_auto} build their model based on commit messages and author identities, while Yan et al. \cite{yan2016automatically} present a Discriminative Probability Latent Semantic Analysis (DPLSA) model for automated categorizing. 
Recently, studies have been adopting new methods to address this problem.
Levin et al. \cite{levin2017boosting} introduce their novel method to predict three types of maintenance tasks. 
Mariano et al. \cite{mariano2019feature} adopt XGBoost, a boosting tree learning algorithm for classification.
Honel et al. \cite{honel2019importance} achieve a high accuracy by adding code density to their prediction model.
Dos et al. \cite{dos2020commit} combine natrual language processing techniques to help train their machine learning model.
Ghadhab et al. \cite{ghadhab2021augmenting} apply deep neutral network classifier and BERT model to predict the categories.

Their models achieve high accuracy, but the categorizations they adopted are too simplistic to clearly characterize all types of commits. 
For example, Levin's model uses only three categories: ``adaptive,'' ``corrective,'' and ``perfective'' while we have 28.

As we want to propose our own more complicated taxonomy, we build our own prediction models based on Random Forest and Extra Tree.

\subsection{Software Quality}

The ultimate goal of this research is to improve software quality.
Either the automation of the classification of commits or commit messages will in the end contribute to improved maintenance process, thus achieving higher software quality.

To evaluate the quality, it is common for researchers and developers to use static analysis tools, such PMD, SonarQube, and FindBugs. 
In addition to the quality metrics provided by these tools, in this research, we also use compilability to evaluate software quality, since the tools only give results when the commits are compilable.
In previous studies on compilability\cite{pooyan_esem, 8170083, Hassan2017ESEM, SMR:SMR1838, pooyan_qrs}, researchers made observations that open-source projects, including popular ones, have uncompilable commits.
By analyzing both compilability and tool-based quality metrics, we investigate how different categories of changes impact the quality and how we can avoid the defects.

\section{Research Questions}
\label{sec:questions}

\subsection{How do different types of commits impact software quality?}
To answer this question, the first problem we have to address is to categorize commits.
In this research, we adopt a taxonomy from a previous study, apply it to the data set, manually categorize 1914 commits, refine the categorization to reduce the ambiguity of the most confusing tag, ``maintenance'', by creating three new sub-categories for ``maintenance''. 
After that, we re-classify the commits with the tag ``maintenance'' and conduct commit pair analysis on software metrics and evaluate how different types of commits impact software compilability.

\subsection{How do we automate classification of commit types and improve performance?}
To make our work more appliable to future new data sets, we automate the prediction of commit types by training multiple models and choose the best one.
And to train the models, we not only use commit messages which is commonly used in previous works, but also use meta-data and quality metrics in our data set.
In this research, we adopt different models for prediction and compare their performances.

\section{Data}
\label{sec:data}

To conduct this purpose-oriented study on commits, we need sufficient data from various open-source projects.
The data set should contain the basic meta-data of projects and software metrics reflecting the quality of those projects.
Thus, we choose to employ SQUAAD, a data set which is collected and used in previous studies \cite{pooyan_esem, pooyan_qrs, qrs2020} as well as our own commit classification data. 

\subsection{SQUAAD Data Set}
The SQUAAD data set includes data from 68 official projects from Apache, Google, and Netflix, each of which contains less than 3000 commits by April 2017.
For project selection, the SQUAAD selects systems that require Maven, Gradle, or Ant for compilation. 
The selected projects do not need extra tools that require manual installation (for example, Protoc) to compile, and they are not Bazel, Eclipse or Android projects. 

The data set provides information of 120731 commits, 39002 out of which are considered as impactful commits.
Impactful commits, as defined in the original data set, are commits that change code in core modules, and a core module is a module that contains the majority of the source code and core functionalities.
The commit information includes not only basic commit data from GitHub, such as the commit times and author email addresses, but also software metrics from PMD, SonarQube, FindBugs, for example, total size and number of packages, as well as quality-specific metrics, including vulnerability and number of bugs. 
These metrics are used in this research for quality analysis. 
Other than these tool-based metrics, we also use ``compilability'', also in the data set, as a quality metric, since the tools only run on compilable commits.

\subsection{Classification Data}
While the data set provides information of the commits, it does not directly indicate the purposes, which we investigate in this research, of those commits. 
Thus, we manually categorize 314 uncompilable commits, hereafter ``breakers'', and 1600 randomly selected compilable commits, hereafter ``neutrals'', by their purposes, hereafter ``types''.
In this process, we adopt and refine the taxonomy for change types, provided by Hindle et al. \cite{Hindle_cate} and refined by Jincheng et al. \cite{qrs2020}, review the code changes and commit messages of all 1914 commits and classify each into one, or more categories, if it has multiple purposes.
To validate the results of classification, we cross-validate our work by ensuring that each commit and its assigned tags are reviewed by at least two researchers.

\section{Approach}
\label{sec:approach}

To answer the proposed research questions, we need to categorize commits, analyze how different types of commits impact software quality and use the categorized commit set to train a prediction model to automate the process, thus making the process more approachable during software development.
That way, we will in the end provide valuable guidelines for software developers to improve the quality.
In this section, we will introduce how we accomplish each of these steps.

\subsection{Manual Classification and Cross Validation}
The fundamental part of this research is a valid set of commits, categorized by their purposes.
Jincheng et al. provides one with 314 ``breakers'' and 600 ``neutrals'' but when we use the set to train our prediction model, the results are not satisfactory.
Thus, we categorize 1914 commits, including 314 ``breakers'' and 1600 ``neutrals''.

To accomplish this, we categorize the commit by reviewing messages and actual code changes, since the commit messages are not always informative, and messages that accompany large commits sometimes leave out the information of minor changes.
In addition, to resolve the problem of change type hierarchy, which means some changes are sub-changes of other larger changes, and the confusion it causes in categorization, we adopt the concept of ``independent change'' from Jincheng et al.'s work. 
In this way, we review all 1914 commits and assign one or more (if one commit contains multiple independent changes) tags to each of them.
For example, commit 5cee2a1\footnote{https://github.com/apache/calcite/commit/5cee2a1} adds piglet, a new component to the software Apache Calcite, with corresponding testing code and build configuration changes.
In this case, we assign this commit ``build'', ``feature add'' (the new code are added to main module, thus not adding a new module), and ``testing'' tags.

Furthermore, to ensure our results are consistent, we cross-validated our classified set with different team members and compare our results with the results of the study conducted by Jincheng et al. \cite{qrs2020} by running the Fishers' Exact Test on results.

\subsection{Refinement of Categorization to Reduce Ambiguity}

As we classify commits, we notice the ambiguity in the definitions of some commit types, especially the tag ``maintenance''.
A commit with a ``maintenance'' tag in this categorization does not mean the commits contribute to a general maintenance task of this software, but to improve, replace existing functionalities, or changing code that have little impact on core features.
For example, code changes in utility functions and getter and setters for classes are categorized as ``maintenance''.
Its vague definition, we believe, is the major reason resulting in the confusion.
Thus, we refined the categorization by dividing ``maintenance'' into three sub-categories.
Three sub-categories are as follows:
\begin{itemize}
    \item \textbf{Replacement}: Replace current functionalities or function calls with new ones or new packages.
    This does not include utility or convenience function changes.
    For example, commit 03e49f9\footnote{https://github.com/google/closure-compiler/commit/03e49f9} replace function TranspilationPasses.addEs6LatePasses with two functions, addEs6LatePassesBeforeNti and addEs6LatePassesAfterNti.
    This change is inside the core feature rather than a convenience function and does not change any core functionality.
    Thus, it is assigned a ``replacement'' tag for this replacement of function call.
    \item \textbf{Modification}: Improve or update current functionality by changing its core logic. 
    This also does not include ``utility'' changes. 
    For example, commit c2059f1\footnote{https://github.com/apache/calcite/commit/c2059f1} adds code that is inside an existing feature (the function name is ``getPredicates'' but it is not a standard getter for a class, so we do not consider it to be a ``utility'' change) and corresponding testing code. 
    Thus, we assign it with a ``modification'' tag as well as a ``testing'' tag.
    \item \textbf{Utility}: Add/Update utility, convenience functions or other simple functions such as standard class getters and setters. Utility changes do not impact core features.
    For example, commit 12bea29\footnote{https://github.com/apache/commons-bcel/commit/12bea29} adds a one-line function which is a standard setter function for class attribute ``index''. 
    Thus, we assign the commit with a ``utility'' tag.
\end{itemize}

After we define these three sub-categories for tag ``maintenance'', we review the commits with ``maintenance'' again to tag them with these new categories.
We also cross-validate the results of this review to make sure that everyone understands the definitions in the same way, and that they do not cause further confusion.

\subsection{Automated Classification}

To automate the classification of commits and train the prediction model, we use the commits that are manually categorized and their commit messages as well as some quality metrics.

However, while the quality metrics are simple and unambiguous, the commit messages are not. 
A piece of commit message may not follow grammar rules, or may contain useless information, such as URLs which won't contribute to the prediction model. 
For example, commit 126e976\footnote{https://github.com/apache/avro/commit/126e976} from Apache Avro provides a message: ``AVRO-906. Java: Fix so that ordering of schema properties is consistent git-svn-id: https://svn.apache.org/repos/asf/avro/trunk@1179356 13f79535-47bb-0310-9956-ffa450edef68.''
It contains an issue ID, a brief description of purpose, and an SVN link as well as ID.
Among them, the issue ID, ``AVRO-906'', and the SVN URL as well as ID are not informative for predicting the purpose of this commit.

Thus, to remove useless information, we first preprocess them by extracting the important information from the messages and remove the noises.
Also, we remove the stop words, which is a set of commonly used words in a language (in our data set, English), and punctuations. 
In addition, we exclude some words from the original commit messages to reduce the size of the vocabulary and further reduce the useless information. 

Following preprocessing, we extract features from messages. 
We convert the messages to high dimension vectors by adopting commonly-used embedding methods in machine learning, including GloVe (Global Vectors for Word Representation), BoW (Bag of Word), and TF-IDF (Term Frequency-Inverse Document Frequency). In these experiments, Bow showed strong interpretability and felicity.
The following code block shows how we configure the prediction model:

\begin{lstlisting}[language=Python]
  # data process
  REMOVE_STOP_WORDS = True
  REMOVE_PUNCTUATION = True
  LEMMATIZE = True
  # BoW feature
  from sklearn.feature_extraction.text import CountVectorizer
  count_vectorizer = CountVectorizer(
    ngram_range=(1, 1), max_features=None,
  )
  count_vectorizer.fit(texts)
  X = count_vectorizer.transform(texts)
  
  # ExtraTree Classifier
  from sklearn.ensemble import ExtraTreesClassifier
  
  clf = ExtraTreesClassifier(
    max_features=1000,
    min_samples_leaf=1,
    max_depth=40,
  )
  preds = clf.predict(X)
  
  \end{lstlisting}
  
  With configuration set, we run the prediction with the Extra Tree classifier from sklearn package:

  \begin{lstlisting}[language = Python]
  from sklearn.ensemble import ExtraTreesClassifier
  X, Y = dataset
  clf = ExtraTreesClassifier()
  preds = clf.predict(X)
  \end{lstlisting}

In addition to commit messages, we also use project meta-data and software metrics in our prediction model to improve the performance, since the data set provides a variety of them and they provide further details of these commits. 
In total, we use 10 quality metrics as additional features, including commit times of the commits, changes of number of classes, files, functions, lines of code, bugs, and sizes between current and its previous commit, as well as the project names and contributors' email. 

Once set up, we train our model again to predict types of the commits by adopting four approaches: SVM (Supported Vector Machines), Decision Tree, Random Forest, and Extra Tree. 
In addition, we conducted a single-label classification which aims to predict one most possible type a commit should be categorized, and a multi-label classification which aims to predict all change types of the commits. 

After we train our prediction model, we also analyze the results, including: 
\begin{itemize}
    \item Analyze the testing tag's influence on the performance. 
    \item Analyze the relative importance of the quality metrics using multi-factor analysis of variance. 
    \item Analyze important features of all tags by constructing the decision-tree of the classifiers. 
\end{itemize}

\subsection{Quality Analysis}

This research starts from categorizing commits, but its ultimate goal has always been improving software quality.
Thus, after we finish the manual classification, we investigate the relations between the commit types and software quality, from two different aspects.
On the one hand, we analyze software metrics provided by PMD, SonarQube and FindBugs, and to make our results more meaningful, we search for the impactful parent commits of these ``neutral'' commits and record whether current commits experience a software metrics increment after their changes are made, compared to their parents.
On the other hand, as we also consider compilability as an important aspect of software quality, and it is not provided by static analysis tools.
Thus, we also analyze how different types of commits impact software compilability.

We conduct statistical significance and correlation analysis on different types of commits to show they do impact software quality. 
After that, we draw conclusions, aiming at warning developers when they made a certain type of change to the repository.

\subsubsection{Tool-based Software Metrics}

To reveal the relation between commit types and software metrics from SonarQube, PMD and FindBugs, we adopt two different approaches, Fisher's Exact Test and Pearson correlation coefficient.

Before running the tests, we group our data into four sub-sets for each commit type, by whether they are assigned a certain type tag or not, and whether they experience an increment of a metric (value ``1'' for an increment, ``0'' for remaining the same or experiencing a reduction), in other words, into contingency tables.

With these tables, we first run Fisher's Exact Test which is used for statistical significance testing.
We apply it to show there is a significant difference of software metrics changes between different types of impactful commits.
That is, to show after certain types of commits and corresponding changes to code, the metrics changes differently.

Furthermore, we analyze correlation between metrics changes and the impactful commit changes by 
calculating Pearson correlation coefficients, which indicate whether there exist strong correlations between certain commit types and metrics or not.

\subsubsection{Compilability}

As we consider compilability as an important aspect of software quality, we also analyze how different types of commits impact software compilability and whether certain types of commits has a higher or lower chance to break compilability.
Using both two-tailed and one-tailed Fisher's Exact Test, we succeed in revealing relations between the commit types and compilability.

\section{Results}
\label{sec:results}

\subsection*{\textbf{RQ1: How do different types of commits impact software quality?}}

\subsubsection{Manual Classification and Validation}
\begin{figure}[htbp]
  \centerline{\includegraphics[scale=1]{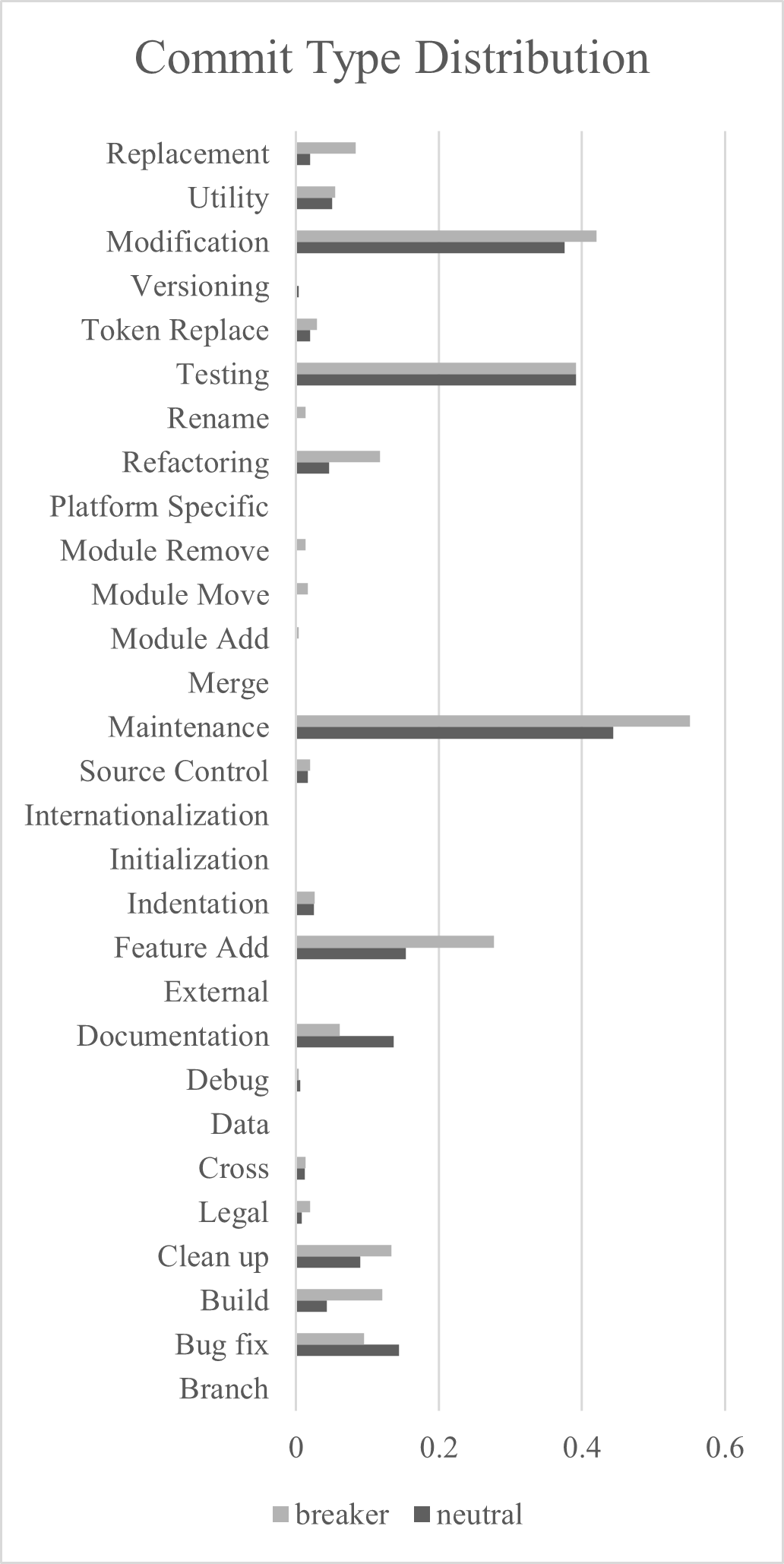}}
  \caption{Commit Type Distribution for Neutrals and Breakers}
  \label{fig:cate}
\end{figure}

To answer this research question we first categorize commits, including 1600 ``neutrals''  and 314 ``breakers''.
Table. \ref{tab:qrs} and Fig. \ref{fig:cate} shows the results of the manual classification.

In Fig. \ref{fig:cate}, light and dark gray bars stand for the percentage of each type of commits that appear in the ``breaker'' set and ``neutral'' set.
For example, the tags, ``feature add'' and ``build'' have higher percentages in breakers while tags ``documentation'' and ``bug fix'' have lower, which we explain as commits that add new features or change build configurations are more likely to cause compilability breach.
Bug fixes and documentation changes, on the contrary, have positive impact on software with respect to reducing the chance of becoming uncompilable.
These conclusions align with our common sense and the results from the previous study\cite{qrs2020}. 

To further validate our results and the categorization for commits, we also compare our results with those from the work of Jincheng et al. which is shown in Table. \ref{tab:qrs}.
The table contains the number of tagged neutral commits for each category.
The reason why we only choose to compare neutrals is that we analyze the same set of ``breakers'', which contains 314 commits, and we consider it to be an insufficient validation to compare the results of ``breakers''.
Thus, we present the results of comparing our neutrals set, consisting of 1600 commits and theirs of 600 commits.
In the table, column ``Tagged-J'' shows the results from Jincheng et al. and the column ``Tagged'' presents ours.
The final column is for the results of the Fisher's Exact Test.
For example, for ``documentation'', in previous study, 81 out of 600 commits are assigned this tag while in our study, 218 out of 1600 are assigned.
The p-value for ``documentation'', which is 1, indicates our results are consistent.

As Table. \ref{tab:qrs} indicates, most types didn't show significant difference except the tag ``bug fix'', and we consider this shows our results are valid.

\begin{table}[htbp]
  \centering
  \caption{Distribution Compared to Previous Study}
    \begin{tabular}{l|rrr}
    \toprule
          & \multicolumn{1}{l}{Tagged-J} & \multicolumn{1}{l}{Tagged} & \multicolumn{1}{l}{p-value} \\
    \midrule
    Branch & 0     & 0     & 1.00 \\
    \midrule
    Bug fix & 109   & 230   & \textbf{0.03} \\
    \midrule
    Build & 22    & 69    & 0.55 \\
    \midrule
    Clean up & 60    & 144   & 0.46 \\
    \midrule
    Legal & 4     & 12    & 1.00 \\
    \midrule
    Cross & 10    & 19    & 0.40 \\
    \midrule
    Data  & 0     & 1     & 1.00 \\
    \midrule
    Debug & 7     & 9     & 0.16 \\
    \midrule
    Documentation & 81    & 218   & 1.00 \\
    \midrule
    External & 0     & 0     & 1.00 \\
    \midrule
    Feature Add & 101   & 245   & 0.39 \\
    \midrule
    Indentation & 21    & 39    & 0.19 \\
    \midrule
    Initialization & 0     & 1     & 1.00 \\
    \midrule
    Internationalization & 1     & 1     & 0.47 \\
    \midrule
    Source Control & 6     & 26    & 0.32 \\
    \midrule
    Maintenance & 251   & 709   & 0.31 \\
    \midrule
    Merge & 1     & 1     & 0.47 \\
    \midrule
    Module Add & 0     & 0     & 1.00 \\
    \midrule
    Module Move & 1     & 1     & 0.47 \\
    \midrule
    Module Remove & 1     & 1     & 0.47 \\
    \midrule
    Platform Specific & 0     & 0     & 1.00 \\
    \midrule
    Refactoring & 26    & 73    & 0.91 \\
    \midrule
    Rename & 0     & 1     & 1.00 \\
    \midrule
    Testing & 221   & 626   & 0.35 \\
    \midrule
    Token Replace & 12    & 32    & 1.00 \\
    \midrule
    Versioning & 2     & 5     & 1.00 \\
    \midrule
    Totally Analyzed & 600   & 1600  &  \\
    \end{tabular}%
  \label{tab:qrs}%
\end{table}%

\subsubsection{Impacts on Software Metrics}
With the manual classification done, we obtain a solid data set for quality analysis.
Firstly, we analyze how software metrics changes on impactful pairs.

We investigate 1600 neutrals we classified and filter their impactful parents, 1578 commits in total, from the data set.
The set of parent commits consists of only 1578 commits instead of 1600 due to various reasons, such as that commits may share the same impactful parent and that some other commits are initial commits of these projects which do not have parents.

We begin with the Fisher's Exact Test (two-tailed, since we do not make ``greater'' or ``less'' assumption in this step) on the commit pairs to find evidence of whether there is a statistically significant difference on software metric increments for each type of changes.
Part of results are shown in Table. \ref{tab:sig}, in which each column stands for a quality metric from FindBugs while each row stands for a commit type.
We only present part of our results here because we have more than 80 metrics and 29 commit types, and it is unnecessary to present all to show evidence supporting our assumption.
As indicated in the table, most entries in the table have a p-value that is less than 0.05, which means there is a significant difference, thus being a piece of evidence of the assumption that certain types of commits may have positive impact on software metrics.

\begin{table*}[htbp]
  \centering
  \caption{The Fisher's Exact Test Results for Part of FindBugs' Metrics}
    \begin{tabular}{p{7.765em}|rrrrrrr}
    \toprule
    \multicolumn{1}{r|}{} & \multicolumn{1}{p{4.135em}}{total\_size} & \multicolumn{1}{p{6.465em}}{num\_packages} & \multicolumn{1}{p{5.235em}}{total\_classes} & \multicolumn{1}{p{4.335em}}{total\_bugs} & \multicolumn{1}{p{3.865em}}{priority\_1} & \multicolumn{1}{p{4.2em}}{priority\_2} & \multicolumn{1}{l}{referenced\_classes} \\
    \midrule
    Branch & \textbf{0.00} & \textbf{0.00} & \textbf{0.00} & \textbf{0.00} & \textbf{0.00} & \textbf{0.00} & \textbf{0.00} \\
    \midrule
    Bug fix & \textbf{0.00} & \textbf{0.00} & 0.38  & \textbf{0.00} & \textbf{0.00} & \textbf{0.00} & 0.40 \\
    \midrule
    Build & \textbf{0.00} & \textbf{0.00} & \textbf{0.00} & \textbf{0.01} & \textbf{0.00} & 0.12  & \textbf{0.00} \\
    \midrule
    Clean up & \textbf{0.00} & \textbf{0.00} & \textbf{0.00} & \textbf{0.01} & \textbf{0.00} & \textbf{0.00} & \textbf{0.00} \\
    \midrule
    Legal & \textbf{0.00} & \textbf{0.05} & \textbf{0.00} & \textbf{0.00} & \textbf{0.00} & \textbf{0.00} & \textbf{0.00} \\
    \midrule
    Cross & \textbf{0.00} & 0.45  & \textbf{0.00} & \textbf{0.00} & 0.09  & \textbf{0.00} & \textbf{0.00} \\
    \midrule
    Data  & \textbf{0.00} & \textbf{0.00} & \textbf{0.00} & \textbf{0.00} & \textbf{0.00} & \textbf{0.00} & \textbf{0.00} \\
    \midrule
    Debug & \textbf{0.00} & \textbf{0.01} & \textbf{0.00} & \textbf{0.00} & \textbf{0.00} & \textbf{0.00} & \textbf{0.00} \\
    \midrule
    Documentation & \textbf{0.00} & \textbf{0.00} & 0.80  & \textbf{0.00} & \textbf{0.00} & \textbf{0.00} & 0.15 \\
    \midrule
    External & \textbf{0.00} & \textbf{0.00} & \textbf{0.00} & \textbf{0.00} & \textbf{0.00} & \textbf{0.00} & \textbf{0.00} \\
    \midrule
    Feature Add & \textbf{0.00} & \textbf{0.00} & 0.11  & \textbf{0.00} & \textbf{0.00} & \textbf{0.00} & 0.92 \\
    \midrule
    Indentation & \textbf{0.00} & 0.10  & \textbf{0.00} & \textbf{0.00} & 0.47  & \textbf{0.00} & \textbf{0.00} \\
    \midrule
    Initialization & \textbf{0.00} & \textbf{0.00} & \textbf{0.00} & \textbf{0.00} & \textbf{0.00} & \textbf{0.00} & \textbf{0.00} \\
    \midrule
    Internationalization & \textbf{0.00} & \textbf{0.00} & \textbf{0.00} & \textbf{0.00} & \textbf{0.00} & \textbf{0.00} & \textbf{0.00} \\
    \midrule
    Source Control & \textbf{0.00} & 1.00  & \textbf{0.00} & \textbf{0.00} & 0.51  & \textbf{0.00} & \textbf{0.00} \\
    \midrule
    Maintenance & \textbf{0.00} & \textbf{0.00} & \textbf{0.00} & \textbf{0.00} & \textbf{0.00} & \textbf{0.00} & \textbf{0.00} \\
    \midrule
    Merge & \textbf{0.00} & \textbf{0.00} & \textbf{0.00} & \textbf{0.00} & \textbf{0.00} & \textbf{0.00} & \textbf{0.00} \\
    \midrule
    Module Add & \textbf{0.00} & \textbf{0.00} & \textbf{0.00} & \textbf{0.00} & \textbf{0.00} & \textbf{0.00} & \textbf{0.00} \\
    \midrule
    Module Move & \textbf{0.00} & \textbf{0.00} & \textbf{0.00} & \textbf{0.00} & \textbf{0.00} & \textbf{0.00} & \textbf{0.00} \\
    \midrule
    Module Remove & \textbf{0.00} & \textbf{0.00} & \textbf{0.00} & \textbf{0.00} & \textbf{0.00} & \textbf{0.00} & \textbf{0.00} \\
    \midrule
    Platform Specific & \textbf{0.00} & \textbf{0.00} & \textbf{0.00} & \textbf{0.00} & \textbf{0.00} & \textbf{0.00} & \textbf{0.00} \\
    \midrule
    Refactoring & \textbf{0.00} & \textbf{0.00} & \textbf{0.00} & \textbf{0.02} & \textbf{0.00} & 0.23  & \textbf{0.00} \\
    \midrule
    Rename & \textbf{0.00} & \textbf{0.00} & \textbf{0.00} & \textbf{0.00} & \textbf{0.00} & \textbf{0.00} & \textbf{0.00} \\
    \midrule
    Testing & \textbf{0.00} & \textbf{0.00} & \textbf{0.00} & \textbf{0.00} & \textbf{0.00} & \textbf{0.00} & \textbf{0.00} \\
    \midrule
    Token Replace & \textbf{0.00} & 0.42  & \textbf{0.00} & \textbf{0.00} & 1.00  & \textbf{0.00} & \textbf{0.00} \\
    \midrule
    Versioning & \textbf{0.00} & \textbf{0.00} & \textbf{0.00} & \textbf{0.00} & \textbf{0.00} & \textbf{0.00} & \textbf{0.00} \\
    \midrule
    Modification & \textbf{0.00} & \textbf{0.00} & \textbf{0.00} & \textbf{0.00} & \textbf{0.00} & \textbf{0.00} & \textbf{0.00} \\
    \midrule
    Utility & \textbf{0.00} & \textbf{0.00} & \textbf{0.00} & 0.08  & \textbf{0.00} & 0.53  & \textbf{0.00} \\
    \midrule
    Replacement & \textbf{0.00} & 0.42  & \textbf{0.00} & \textbf{0.00} & 1.00  & \textbf{0.00} & \textbf{0.00} \\
    \bottomrule
    \end{tabular}%
  \label{tab:sig}%
\end{table*}%

After the exact test shows significance for most entries, we apply Pearson's further reveal the potential linear correlation between certain type of change and any metrics by running the test on all types and metrics.
The reason why we analyze all metrics in this step is that although some metrics are more quality-focused, such as vulnerability and security, but we believe other metrics, such as lines of code and number of classes, also, to some extent, represents aspects of software quality, which we prefer keeping track of to completely leaving them out.

The results of the analysis on 1578 pairs are presented in following tables.
Table. \ref{tab:pmd}, Table. \ref{tab:sq}, and Table. \ref{tab:fb} respectively show how software metrics change (whether they experience an increment or not) for 1578 ``neutrals'' from their impactful parents for PMD, SonarQube and FindBugs.
To improve the readability of results, we select columns, each of which stands for a commit type, that show statistically significant quality metrics changes between parent commits and child commits. 
Column ``CLN'' stands for ``clean up'', ``DOC'' for ``documentation'', ``FTA'' for ``feature add'', ``TST'' for ``testing'', and ``UTL'' for ``utility'' while rows stand for software metrics.
Entries in \textbf{bold} are those with correlation coefficients whose absolute value is larger than 0.2 (we mark by this rule only with the intention to emphasize values that are relatively high).

\begin{table}[htbp]
    \centering
    \caption{PMD}
      \begin{tabular}{p{10.465em}|ccccc}
      \toprule
      \multicolumn{1}{c|}{} & \multicolumn{1}{p{2.1em}}{CLN} & \multicolumn{1}{p{2.1em}}{DOC} & \multicolumn{1}{p{2.1em}}{FTA} & \multicolumn{1}{p{2.1em}}{TST} & \multicolumn{1}{p{2.1em}}{UTL} \\
      \midrule
      basic & -0.03 & -0.04 & 0.16  & 0.12  & -0.01 \\
      \midrule
      emptycode & 0.02  & -0.06 & 0.09  & 0.08  & 0.03 \\
      \midrule
      cloneimplementation & -0.01 & -0.01 & 0.08  & 0.04  & -0.01 \\
      \midrule
      comments & -0.16 & -0.10 & \textbf{0.45} & \textbf{0.27} & 0.12 \\
      \midrule
      codesize & -0.07 & -0.13 & \textbf{0.36} & \textbf{0.21} & 0.04 \\
      \midrule
      stringandstringbuffer & -0.01 & -0.04 & 0.19  & 0.09  & 0.02 \\
      \midrule
      naming & -0.10 & -0.16 & \textbf{0.47} & \textbf{0.26} & 0.04 \\
      \midrule
      strictexceptions & -0.06 & -0.07 & \textbf{0.24} & 0.11  & -0.01 \\
      \midrule
      optimization & -0.12 & \textbf{-0.24} & \textbf{0.49} & \textbf{0.32} & 0.14 \\
      \midrule
      design & -0.09 & -0.16 & \textbf{0.43} & \textbf{0.23} & 0.04 \\
      \midrule
      securitycodeguidelines & 0.00  & -0.03 & 0.12  & 0.05  & 0.02 \\
      \midrule
      braces & -0.04 & -0.08 & 0.14  & 0.08  & 0.00 \\
      \midrule
      typeresolution & -0.01 & -0.07 & \textbf{0.22} & 0.10  & -0.01 \\
      \midrule
      coupling & -0.12 & \textbf{-0.24} & \textbf{0.39} & \textbf{0.31} & 0.01 \\
      \midrule
      importstatements & 0.01  & -0.08 & 0.17  & 0.13  & 0.00 \\
      \midrule
      unusedcode & -0.05 & -0.05 & 0.15  & 0.13  & 0.02 \\
      \midrule
      unnecessary & -0.02 & -0.09 & \textbf{0.25} & 0.18  & 0.02 \\
      \bottomrule
      \end{tabular}%
    \label{tab:pmd}%
\end{table}%

\begin{table}[htbp]
    \centering
    \caption{SonarQube}
      \begin{tabular}{p{11.5em}|ccccc}
      \toprule
      \multicolumn{1}{c|}{} & \multicolumn{1}{p{2.1em}}{CLN} & \multicolumn{1}{p{2.1em}}{DOC} & \multicolumn{1}{p{2.1em}}{FTA} & \multicolumn{1}{p{2.1em}}{TST} & \multicolumn{1}{p{2.1em}}{UTL} \\
      \midrule
      total & -0.13 & \textbf{-0.22} & \textbf{0.40} & \textbf{0.30} & 0.04 \\
      \midrule
      info  & -0.03 & 0.01  & 0.14  & 0.08  & 0.01 \\
      \midrule
      minor & -0.03 & -0.15 & \textbf{0.36} & 0.18  & 0.01 \\
      \midrule
      major & -0.13 & \textbf{-0.23} & \textbf{0.42} & \textbf{0.32} & 0.04 \\
      \midrule
      critical & -0.05 & -0.10 & \textbf{0.29} & 0.13  & 0.02 \\
      \midrule
      blocker & -0.02 & -0.04 & 0.15  & 0.10  & 0.00 \\
      \midrule
      codesmell & -0.14 & \textbf{-0.22} & \textbf{0.40} & \textbf{0.30} & 0.04 \\
      \midrule
      bug   & 0.00  & -0.07 & \textbf{0.20} & 0.12  & 0.00 \\
      \midrule
      vulnerability & -0.05 & -0.06 & 0.18  & 0.09  & 0.01 \\
      \midrule
      infocodesmell & -0.03 & 0.01  & 0.14  & 0.08  & 0.01 \\
      \midrule
      minorcodesmell & -0.03 & -0.15 & \textbf{0.36} & 0.18  & 0.01 \\
      \midrule
      majorcodesmell & -0.12 & \textbf{-0.23} & \textbf{0.42} & \textbf{0.31} & 0.04 \\
      \midrule
      majorbug & 0.03  & -0.02 & 0.12  & 0.04  & -0.01 \\
      \midrule
      criticalcodesmell & -0.04 & -0.08 & \textbf{0.24} & 0.08  & 0.02 \\
      \midrule
      criticalbug & 0.01  & -0.05 & 0.16  & 0.08  & -0.01 \\
      \midrule
      criticalvulnerability & -0.05 & -0.06 & 0.17  & 0.08  & 0.02 \\
      \midrule
      blockerbug & -0.03 & -0.04 & 0.13  & 0.10  & 0.01 \\
      \midrule
      blockervulnerability & -0.01 & -0.02 & 0.10  & 0.07  & -0.02 \\
      \midrule
      classes & -0.06 & -0.13 & \textbf{0.58} & \textbf{0.27} & -0.01 \\
      \midrule
      comment\_lines\_density & 0.07  & 0.06  & 0.12  & 0.10  & 0.02 \\
      \midrule
      vulnerabilities & -0.05 & -0.06 & 0.18  & 0.09  & 0.01 \\
      \midrule
      lines & \textbf{-0.30} & -0.07 & \textbf{0.29} & \textbf{0.26} & 0.15 \\
      \midrule
      ncloc & \textbf{-0.27} & \textbf{-0.25} & \textbf{0.35} & \textbf{0.31} & 0.17 \\
      \midrule
      complexity & \textbf{-0.21} & \textbf{-0.31} & \textbf{0.43} & \textbf{0.36} & \textbf{0.23} \\
      \midrule
      major\_violations & -0.11 & -0.19 & \textbf{0.48} & \textbf{0.31} & 0.03 \\
      \midrule
      duplicated\_blocks & -0.01 & -0.07 & 0.19  & 0.10  & 0.00 \\
      \midrule
      code\_smells & -0.14 & -0.18 & \textbf{0.43} & \textbf{0.28} & 0.04 \\
      \midrule
      file\_complexity & 0.00  & -0.14 & \textbf{0.23} & \textbf{0.21} & 0.11 \\
      \midrule
      functions & -0.12 & \textbf{-0.22} & \textbf{0.55} & \textbf{0.37} & \textbf{0.28} \\
      \midrule
      duplicated\_files & -0.01 & -0.06 & \textbf{0.22} & 0.08  & -0.03 \\
      \midrule
      violations & -0.05 & -0.10 & \textbf{0.29} & 0.13  & 0.02 \\
      \midrule
      majorbug & -0.14 & -0.18 & \textbf{0.42} & \textbf{0.28} & 0.04 \\
      \midrule
      statements & \textbf{-0.22} & \textbf{-0.33} & \textbf{0.41} & \textbf{0.36} & \textbf{0.20} \\
      \midrule
      blocker\_violations & -0.02 & -0.04 & 0.15  & 0.09  & 0.00 \\
      \midrule
      reliability\_remediation\_effort & 0.00  & -0.07 & \textbf{0.20} & 0.12  & 0.00 \\
      \midrule
      duplicated\_lines & -0.03 & -0.04 & 0.17  & 0.09  & -0.01 \\
      \midrule
      bugs  & 0.00  & -0.07 & \textbf{0.21} & 0.11  & 0.00 \\
      \midrule
      security\_remediation\_effort & -0.05 & -0.06 & 0.18  & 0.09  & 0.01 \\
      \midrule
      directories & -0.04 & -0.04 & \textbf{0.20} & 0.09  & -0.03 \\
      \midrule
      info\_violations & -0.04 & -0.03 & \textbf{0.21} & 0.14  & -0.01 \\
      \midrule
      sqale\_index & -0.14 & \textbf{-0.22} & \textbf{0.40} & \textbf{0.30} & 0.04 \\
      \midrule
      minor\_violations & -0.04 & -0.14 & \textbf{0.36} & 0.17  & 0.00 \\
      \midrule
      files & -0.05 & -0.09 & \textbf{0.53} & \textbf{0.23} & -0.04 \\
      \bottomrule
      \end{tabular}%
    \label{tab:sq}%
\end{table}%

\begin{table}[htbp]
    \centering
    \caption{FindBugs}
      \begin{tabular}{p{8.335em}|rrrrr}
      \toprule
      \multicolumn{1}{r|}{} & \multicolumn{1}{p{2.1em}}{CLN} & \multicolumn{1}{p{2.1em}}{DOC} & \multicolumn{1}{p{2.1em}}{FTA} & \multicolumn{1}{p{2.1em}}{TST} & \multicolumn{1}{p{2.1em}}{UTL} \\
      \midrule
      total\_size & \textbf{-0.25} & \textbf{-0.32} & \textbf{0.36} & \textbf{0.33} & 0.18 \\
      \midrule
      num\_packages & -0.02 & -0.04 & \textbf{0.20} & 0.12  & -0.03 \\
      \midrule
      total\_classes & -0.06 & -0.14 & \textbf{0.56} & \textbf{0.29} & 0.00 \\
      \midrule
      total\_bugs & -0.02 & -0.10 & \textbf{0.26} & 0.12  & -0.01 \\
      \midrule
      priority\_1 & -0.03 & -0.06 & 0.16  & 0.09  & 0.01 \\
      \midrule
      priority\_2 & -0.02 & -0.09 & \textbf{0.25} & 0.11  & -0.03 \\
      \midrule
      referenced\_classes & -0.05 & -0.15 & \textbf{0.52} & \textbf{0.30} & 0.01 \\
      \midrule
      bad\_practice & -0.01 & -0.05 & 0.13  & 0.06  & -0.01 \\
      \midrule
      malicious\_code & -0.02 & -0.05 & 0.16  & 0.08  & 0.00 \\
      \midrule
      performance & -0.01 & -0.04 & 0.09  & 0.05  & -0.02 \\
      \midrule
      correctness & 0.01  & -0.04 & 0.11  & 0.07  & 0.01 \\
      \midrule
      style & 0.02  & -0.06 & \textbf{0.20} & 0.10  & -0.02 \\
      \midrule
      experimental & -0.01 & -0.02 & 0.02  & 0.05  & -0.01 \\
      \midrule
      mt\_corectness & -0.03 & -0.03 & 0.08  & 0.03  & -0.02 \\
      \midrule
      i18n  & -0.03 & -0.03 & 0.14  & 0.03  & -0.02 \\
      \bottomrule
      \end{tabular}%
    \label{tab:fb}%
\end{table}%

As indicated in Table. \ref{tab:pmd}, Table. \ref{tab:sq}, and Table. \ref{tab:fb}, commits with tag ``feature add'' show relatively strong correlations with many software metrics, including size-focused ones, such as ``codesize'' of PMD, ``ncloc'' of SonarQube and ``total\_size'' of FindBugs, and quality-focused ones, such as ``majorbug'', ``major\_violations'' of SonarQube and ``total\_bugs'' of FindBugs.
Commits with tags ``documentation'' and ``testing'' show correlations with some metrics, such as ``complexity'' and ``codesmell'' of SonarQube while others with tags ``clean up'' and ``utility'' only show weak correlations to a few metrics such as ``statements'' of SonarQube.

The given three tables indicate that when certain types of changes are made by the developers to the software, some aspects of software quality changes correspondingly, especially significantly when it is adding a new feature.

\subsubsection{Impacts on Compilability}

In addition to software metrics, we also perform the Fisher's Exact Test on ``breakers'' and ``neutrals''. 
Instead of what we do with respect to metrics, we not only run ``two-tailed'' tests for software metrics, but also apply ``one-tailed'' tests (all of them are provided by python package ``scipy.stats'') for breakers to reveal whether certain types of changes have positive or negative impacts on software compilability.

The results are shown in Table. \ref{tab:compilability}.
Tags ``bug fix'' and ``documentation'' show statistical significance that indicate they tend to reduce the chance of commits to break while Tags ``build'', ``clean up'', ``feature add'', ``maintenance'', ``module move'', ``module remove'', ``refactoring'', ``rename'', and ``replacement'' tend to increase the chance of breaking the compilability.

These conclusions can serve as guidelines and warn developers when they push specific types of commits, as mentioned above, to the software repositories.

\begin{table}[htbp]
  \centering
  \caption{Impacts of Different Types of Commits on Compilability}
    \begin{tabular}{p{9em}|ccc}
    \toprule
    \multicolumn{1}{c|}{\multirow{2}[4]{*}{}} & \multicolumn{3}{c}{P-value} \\
\cmidrule{2-4}    \multicolumn{1}{c|}{} & \multicolumn{1}{p{5em}|}{\centering{two-sided}} & \multicolumn{1}{p{5em}|}{\centering{greater}} & \multicolumn{1}{p{5em}}{\centering{less}} \\
    \midrule
    Branch & 1.00  & 1.00  & 1.00 \\
    \midrule
    Bug fix & \textbf{0.02} & \textbf{0.01} & 0.99 \\
    \midrule
    Build & \textbf{0.00} & 1.00  & \textbf{0.00} \\
    \midrule
    Clean up & \textbf{0.02} & 0.99  & \textbf{0.01} \\
    \midrule
    Legal & 0.10  & 0.98  & 0.06 \\
    \midrule
    Cross & 0.78  & 0.68  & 0.54 \\
    \midrule
    Data  & 1.00  & 0.84  & 1.00 \\
    \midrule
    Debug & 1.00  & 0.49  & 0.83 \\
    \midrule
    Documentation & \textbf{0.00} & \textbf{0.00} & 1.00 \\
    \midrule
    External & 1.00  & 1.00  & 1.00 \\
    \midrule
    Feature Add & \textbf{0.00} & 1.00  & \textbf{0.00} \\
    \midrule
    Indentation & 0.84  & 0.64  & 0.52 \\
    \midrule
    Initialization & 1.00  & 0.84  & 1.00 \\
    \midrule
    Internationalization & 1.00  & 0.84  & 1.00 \\
    \midrule
    Source Control & 0.64  & 0.74  & 0.43 \\
    \midrule
    Maintenance & \textbf{0.00} & 1.00  & \textbf{0.00} \\
    \midrule
    Merge & 1.00  & 0.84  & 1.00 \\
    \midrule
    Module Add & 0.16  & 1.00  & 0.16 \\
    \midrule
    Module Move & \textbf{0.00} & 1.00  & \textbf{0.00} \\
    \midrule
    Module Remove & \textbf{0.00} & 1.00  & \textbf{0.00} \\
    \midrule
    Platform Specific & 1.00  & 1.00  & 1.00 \\
    \midrule
    Refactoring & \textbf{0.00} & 1.00  & \textbf{0.00} \\
    \midrule
    Rename & \textbf{0.00} & 1.00  & \textbf{0.00} \\
    \midrule
    Testing & 1.00  & 0.53  & 0.52 \\
    \midrule
    Token Replace & 0.39  & 0.88  & 0.22 \\
    \midrule
    Versioning & 1.00  & 0.41  & 1.00 \\
    \midrule
    Modification & 0.14  & 0.94  & 0.08 \\
    \midrule
    Utility & 0.78  & 0.68  & 0.42 \\
    \midrule
    Replacement & \textbf{0.00} & 1.00  & \textbf{0.00} \\
    \bottomrule
    \end{tabular}%
  \label{tab:compilability}%
\end{table}%

\subsection*{\textbf{RQ2: How do we automate classification of commit types and improve performance?}}

\begin{figure}[htbp]
  \centerline{\includegraphics[scale=0.18]{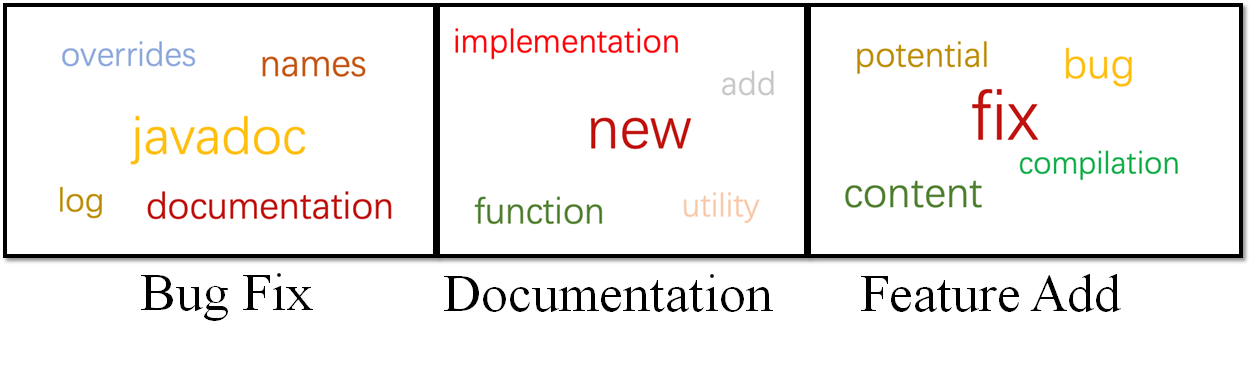}}
  \caption{Keywords for Commit Types}
  \label{fig:feature}
\end{figure}

To predict commit types, we first use commit messages and build a prediction model based on extra trees.
However, since the performance is not satisfying, we add meta-data and quality metrics to train the model, and the first row of Table. \ref{tab:predict} shows the results.
As a side product of this prediction, we also collect keywords from commit messages for different types, part of which are shown in Fig. \ref{fig:feature}.

In addition to adding new information to the prediction model, we also adopt random forest and extra-tree-based multi-label classification models for this task and the results are presented in Table. \ref{tab:comparison}.

Adding new information slightly improve the performance, as shown in the Table. \ref{tab:predict}, but newly adopted models does not perform well. 
Thus, after we refine the categorization and add three sub-categories for tag ``maintenance'', we re-train the prediction model with three different settings: 29 tags (with three new tags), 28 tags (without ``maintenance''), and 25 tags (without ``maintenance'' or three new tags).
The results are also presented in Table. \ref{tab:predict} from the second row to the last row.

By comparing the first two rows or last two rows of Table. \ref{tab:predict}, the performance indicates that despite our attempt to reduce the ambiguity of definitions for the tag ``maintenance'', the prediction model accuracy is not improved.
But the tag ``maintenance'' obviously impacts the prediction model negatively, since when we remove it, the performance improves. 

Overall, these prediction models does not show a high prediction accuracy, but as we are the first one to use as many as 28 types to train the prediction model for commit change types, it is reasonable.

\begin{table*}[htbp]
  \centering
  \caption{Prediction Model Performance}
    \begin{tabular}{p{11.465em}|ccc|ccc}
    \toprule
    \multicolumn{1}{c|}{\multirow{2}[4]{*}{}} & \multicolumn{3}{c|}{Only Using Commit Message} & \multicolumn{3}{c}{Adding Meta-data and Metrics} \\
\cmidrule{2-7}    \multicolumn{1}{c|}{} & \multicolumn{1}{p{4.1em}}{\centering{acc}} & \multicolumn{1}{p{4.1em}}{\centering{recall}} & \multicolumn{1}{p{4.1em}|}{\centering{f1}} & \multicolumn{1}{p{4.1em}}{\centering{acc}} & \multicolumn{1}{p{4.1em}}{\centering{recall}} & \multicolumn{1}{p{4.1em}}{\centering{f1}} \\
    \midrule
    Original setting (26 categories) & 0.36  & 0.38  & 0.45  & 0.45  & 0.48  & 0.55 \\
    \midrule
    Add 3 new tags (29 categories) & 0.32  & 0.38  & 0.43  & 0.42  & 0.5   & 0.55 \\
    \midrule
    Add 3 new tags and remove maintenance (28 categories) & 0.3   & 0.33  & 0.4   & 0.41  & 0.44  & 0.53 \\
    \midrule
    Remove maintenance (25 categories) & 0.48  & 0.49  & 0.56  & 0.57  & 0.57  & 0.66 \\
    \bottomrule
    \end{tabular}%
  \label{tab:predict}%
\end{table*}%

\begin{table}[htbp]
  \centering
  \caption{Prediction Accuracy Comparison}
    \begin{tabular}{|p{11.465em}|rrr}
    \toprule
    \multicolumn{1}{|r|}{} & \multicolumn{1}{p{4.1em}}{acc} & \multicolumn{1}{p{4.1em}}{recall} & \multicolumn{1}{p{4.1em}}{f1} \\
    \midrule
    Random forest & 0.45  & 0.47  & 0.55 \\
    \midrule
    ExtraTrees & 0.45  & 0.48  & 0.55 \\
    \midrule
    Multi-clf using ExtraTrees & 0.27  & 0.31  & 0.42 \\
    \bottomrule
    \end{tabular}%
  \label{tab:comparison}%
\end{table}%

\section{Threats to Validity}
\label{sec:threats}

This section discusses threats to the validity of this research based on the guidelines created by Wieringa et al. \cite{threats}.

\textbf{External Validity.}
The major threat is our subject data.
We adopt the data set from the SQUAAD data set, which is limited to open-source java projects.
However, the data set contains 68 projects from both for-profit and non-profit organizations, which means it has a fine generalizability.
Still, to generalize our conclusions, it is necessary to investigate software systems from different organizations, developed under different guidelines and process models.

\textbf{Conclusion Validity.}
The major threat to the conclusion validity is the potential mistakes in the manual tasks, including the classification for commit types and code for analyses where human errors are almost unavoidable.
To mitigate it, we assign at least two team members to each task, thus keeping each piece of work cross-validated. 
For example, in the manual classification, each commit is reviewed by at least two researchers to avoid errors.

\textbf{Internal Validity.}
Adopting ambiguous methods, such as an ambiguous taxonomy for commits are major threats to internal validity.
For example, the ``maintenance'' tag in this research is a significant issue that cause confusion.
To resolve it, we create sub-categories for it, review the commits and rule out alternative explanations for those commits and the analysis results.

\textbf{Construct Validity.}
The main threats to construct validity are the validity of measures we apply in this research, including the statistical analysis methods, the prediction models and the taxonomy we adopt for commit classification.
To mitigate them, we tried different methods, compare them, reviewing the documentations, and select the most appropriate one for this research.
For example, we tried different prediction models, compare them and select the best-performance, and review the details of the model construction to confirm it is a correct method for this research.

\section{Conclusions}
\label{sec:conclusion}

This study focuses on categorizing commit by their purposes and investigating how different types of commit impact different aspects of software quality.

We first refined the taxonomy for commits by reducing the ambiguity of a specific category, ``maintenance'', and create sub-categories for it.
In addition, we test a variety of prediction models for the taxonomy on our data set and make attempts to improve it, and both the data set and models are available for future research.

Having classified the commits, we analyze the relations between commit types and software quality, including software metrics and compilability. 
The results indicate that new features in software are most likely to cause various software metrics to change, followed by ``documentation'', ``testing'', ``clean up'', and ``utility'' while ``build'', ``clean up'', ``feature add'', ``maintenance'', ``module move'', ``module remove'', ``refactoring'', ``rename'', and ``replacement'' are more likely to cause compilability breach.
Combined with the prediction model, we will be able to construct a framework to detect and categorize changes made by developers, and warn them when they make certain types of changes that have high risks of introducing defects to the projects.

One of the future steps of this research is to improve the prediction model to serve developers when they contribute to software.
To improve the model, we need to try different models as well as refine the taxonomy.
Another potential future step is investigating further into software quality and study how software evolves or decays as different kinds of efforts accumulate. 
To achieve this, we may adopt other tools, such as CAST for architecture analysis and Tetrad for causality analysis.
In addition, to generalize the conclusions and guidelines, it is necessary to collect additional data from projects that use different languages or that are close-source.

\section*{Acknowledgment}
This material is based upon work supported in part by the U.S. Department of Defense through the Systems Engineering Research Center (SERC) under Contract No. HQ0034-13-D-0004 Research Task WRT 1016 --- ``Reducing Total Ownership Cost (TOC) and Schedule.'' SERC is a federally funded University Affiliated Research Center managed by Stevens Institute of Technology. 

\medskip
\bibliographystyle{./bibliography/IEEEtran}
\bibliography{./bibliography/IEEEabrv,./bibliography/IEEEexample,./bibliography/references}

\vspace{12pt}
\color{red}

\end{document}